\documentclass[aps, prd, twocolumn, showpacs, superscriptaddress, groupedaddress]{revtex4-1} 
\usepackage{graphicx}	
\usepackage{amssymb}
\usepackage{dcolumn}
\usepackage{color}
\usepackage{subfigure, rotating, bm, array}
\usepackage[pagebackref=false, colorlinks=true]{hyperref}
\hypersetup{
linkcolor=blue,     % color of internal links
citecolor=blue,     % color of links to bibliography
urlcolor=blue} 
\begin{document}
\title{Shadow of nulllike and timelike naked singularities without photon spheres}
\author{Dipanjan Dey}
\email{dipanjandey.icc@charusat.ac.in}
\affiliation{International Center for Cosmology, Charusat University, Anand 388421, Gujarat, India}
\author{Rajibul Shaikh}
\email{rshaikh@iitk.ac.in}
\affiliation{Indian Institute of Technology, Kanpur 208016, India}
\author{Pankaj S. Joshi}
\email{psjprovost@charusat.ac.in}
\affiliation{International Center for Cosmology, Charusat University, Anand 388421, Gujarat, India}

\date{\today}

\begin{abstract}
In this paper, we derive general conditions for a shadow to occur, without a photon sphere in a spacetime, caused by central nulllike or timelike naked singularities. Using these conditions, 
we propose classes of spacetimes which have nulllike and timelike naked singularities at the center, and that fulfill these `shadow without photon sphere' conditions. Considering additional asymptotically flat conditions, we show that, for a fixed Schwarzschild mass, the timelike naked singularities can cast a shadow of a size that is equal to, greater or smaller, than the size of a black hole shadow. On the other hand, the size of  shadow of a nulllike naked singularity is always less than that of a black hole. Such novel features of shadows of nulllike and timelike naked singularities in the absence of photon spheres may help us to distinguish between black holes and naked singularities observationally.
\end{abstract}

\maketitle

\section{Introduction}
Recent observations of the shadow of Messier 87 (M87) galactic center by the Event Horizon Telescope (EHT) group \cite{Akiyama:2019fyp}, and the observed stellar motions (i.e. the motion of S02, S102, S46, etc. stars) around our galaxy center (Sgr-A*) by the GRAVITY, SINFONI collaborations \cite{M87, Eisenhauer:2005cv, center1} have triggered a great interest to investigate the causal structure of spacetime around the galactic center. It is expected that the central regions 
of galaxies harbour a supermassive blackhole, and a spacetime singularity within, formed from catastrophic continual gravitational collapse of primordial matter cloud. However, there exists a big mystery regarding the causal structure of the singularity. In general relativity, there are three types of strong spacetime singularities: spacelike, timelike and nulllike. The spacelike singularity is causally disconnected from the other points of spacetime manifold, and they might be hidden within horizons, whereas the nulllike and timelike singularities are causally connected to the other points of spacetime. There are a lot of literature \cite{Joshi:1993zg, JMN11, Mosani1, Mosani2, Dafermos:2017dbw, Dey:2019fja} where it is shown that nulllike and timelike singularities can be formed during the gravitational collapse of physically realistic matter clouds, which contradicts  the Cosmic Censorship Conjecture (CCC), proposed by Rodger Penrose \cite{Penrose}.  

If nulllike and timelike naked singularities do exist in reality and if they are stable under perturbations, then there must be some distinguishable physical signatures of the same. There are many papers where investigation in this direction has been done  \cite{Shaikh:2019hbm,Gralla:2019xty,Abdikamalov:2019ztb,Yan:2019etp,Vagnozzi:2019apd,Gyulchev:2019tvk,Shaikh:2019fpu,Dey:2013yga,Dey+15,Martinez:2019nor,Eva,Eva1,Eva2,tsirulev,Bambhaniya:2019pbr,Joshi:2019rdo,Shaikh:2018lcc,Bambh,Dey:2019fpv,Bhattacharya:2017chr,Joshi2020,Bam2020,Paul2020,Dey:2020haf}. In \cite{Bambhaniya:2019pbr,Dey:2019fpv, Bam2020}, it is shown that the motion of particles around a naked singularity can have a distinguishable signature which cannot be seen around a Schwarzschild black hole and a Kerr black hole. It is shown that a particle moving around a naked singularity can precess in the opposite direction of its motion, which is not allowed in Schwarzschild or in Kerr spacetime. In \cite{Bam2020}, we predict the possible future trajectory of S02 star around Sgr-A* if  it precesses in that way.

Nulllike or timelike naked singularities can also have distinguishable physical signature in the context of gravitational lensing and shadows they might cast. In \cite{Shaikh:2018lcc,Dey:2020haf, Joshi2020, Paul2020}, the gravitational lensing and shadow phenomena in the presence of nulllike and timelike naked singularities are investigated. From these investigations, it is now established that the shadow is not the signature of a black hole alone, it can also be cast by timelike or nulllike naked singularities in the presence of a photon sphere. In \cite{Dey:2020haf, Joshi2020}, it is shown that even a spacetime having a strong central nulllike naked singularity does not need a photon sphere in order to cast shadow. In \cite{Dey:2020haf}, we construct a spacetime configuration which has visible strong singularity at the center and has no photon sphere. We show that this singular spacetime configuration can cast shadow in presence of a thin shell of matter. In \cite{Joshi2020}, we show further that the a nulllike strong naked singularity can itself cast a shadow in the absence of both a photon sphere and a thin shell of matter. The important understanding one can get from all these works is that black holes are no longer considered as the only candidates which can cast shadows. Visible or naked singularity also can cast a shadow which might be distinguishable from a black hole shadow. These results are quite intriguing in the context of the recent and upcoming observational results of EHT collaboration.

In \cite{Paul2020}, it is shown that the singular spacetime proposed in \cite{Joshi2020} has a null singularity at the center. As stated above, this null singularity casts shadow in the absence of a photon sphere. The interesting fact that emerges is, the size of the shadow cast by this nulllike naked singularity is less than the size of shadow of a Schwarzschild black hole, where the ADM mass of the proposed spacetime is  equal to the Schwarzschild mass of the black hole \cite{Joshi2020}. Now, one can ask whether the existence of a shadow without a photon sphere implies the existence of a nulllike naked singularity at the center, or whether a timelike naked singularity can also cast shadow in the absence of a photon sphere. In this paper, we address this issue and predict what would be the nature of the shadow of a timelike naked singularity in the absence of a photon sphere. 

This paper is organized as follows. In Sec.~(\ref{seccond}), we derive the general conditions for the existence of shadow without photon sphere. Using those conditions, we propose a class of spacetimes which have nulllike strong naked singularity at the center and these fulfil the `shadow without photon sphere' conditions. In Sec.~(\ref{sec3}), we derive general conditions for which timelike naked singularities can cast shadow in the absence of a photon sphere. Using these conditions, we propose a class of spacetimes which posses a timelike naked singularity at the center. The timelike singular spacetimes are not asymptotically Minkowskian. Therefore, in Sec.~(\ref{sec4}), we construct an asymptotically flat spacetime configuration which has a timelike naked singularity at the center. We study the shadow which can be cast by this spacetime configuration and compare it with the shadow of a nulllike naked singularity, and that of a Schwarzschild black hole. In Sec.~(\ref{sec5}), we conclude by discussing the important results of this paper. Here, we consider Newton's gravitational constant $G_N$ and light velocity $c$ as unity.

\section{Conditions for shadow without photon sphere}
\label{seccond}
The line element of a spherically symmetric, static spacetime can be written as,
\begin{equation}
dS^2=-A(r)dt^2+B(r)dr^2+r^2d\Omega^2\,\, ,
\label{genspt}
\end{equation}
where $d\Omega^2=d\theta^2+\sin^2\theta~d\phi^2$. The conserved quantities for a particle along a geodesic in the static, spherically spacetime are, $l=r^2\dot{\phi}$ and $e=A(r)\dot{t}$, where $l,e$ are the conserved angular momentum and energy respectively and an overdot represents a differentiation with respect to affine parameter.   For null geodesics, we can write,
\begin{equation}
-\frac{e^2}{A(r)}+ \frac{l^2}{r^2}+B(r)\dot{r}^2=0\,\, ,
\end{equation}
which implies,
\begin{equation}
V_{eff}(r)+A(r)B(r)\dot{r}^2=e^2\,\, ,
\end{equation}
where $V_{eff}(r)=l^2\frac{A(r)}{r^2}$. Absorbing the constant of motion ($l$) inside $A(r)$, we can write down the effective potential of nulllike geodesics as,
\begin{equation}
V_{eff}(r)=\frac{A(r)}{r^2}\,\, .
\end{equation}  
We know that null geodesics have unstable circular orbits when there exists a photon sphere at a particular radius $r_{ph}$ where $V_{eff}(r_{ph})=e^2, V^{\prime}_{eff}(r_{ph})=0 $ and $V^{\prime\prime}_{eff}(r_{ph})<0 $. Therefore, we have to solve the following equation to get the position of a photon sphere,
\begin{equation}
r_{ph}=\frac{2A(r_{ph})}{A^{\prime}(r_{ph})}\,\, ,
\label{rph}
\end{equation} 
where, at $r=r_{ph}$, $\frac{1}{r_{ph}}\left(\frac{2A(r_{ph})}{A^{\prime\prime}(r_{ph})}\right)>1$. 
If the spacetime mentioned in Eq.~(\ref{genspt}) does not allow any photon sphere then one would not find any real, positive solution of Eq.~(\ref{rph}) for $r_{ph}$. It is to be noted that the photon sphere corresponds to the maximum of $V_{eff}$ and $V_{eff}$ vanishes in the limit $r\to \infty$. Therefore, to ensure that no photon sphere exists, $V_{eff}$ must be positive and monotonically decreasing in $r$, i.e., we must have  $V_{eff}^{\prime}(r)<0~\forall~r$. Now, along with this no photon sphere condition, if we impose another condition that $V_{eff}$ has a finite positive value  at $r=0$, then $\lim_{r \to 0}\frac{A(r)}{r^2}$ should be finite. As we know, the impact parameter $b=\frac{l}{e}=\frac{r_{tp}}{\sqrt{A(r_{tp})}}$, where $r_{tp}$ is the radius of the turning point of the nulllike geodesic. Since we demand finite positive value of $V_{eff}$ at $r=0$, there is a critical value of the impact parameter for which the turning point becomes $r_{tp}=0$. Any ingoing geodesic having impact parameter greater than the critical value will take a turn at a radius outside the singularity and escape to a faraway observer. On the other hand, geodesic with impact parameter less than the critical value will be terminated at $r=0$, and as a result, an  asymptotic static observer should see a shadow of radius equal to the critical impact parameter. Therefore, though there exist no photon sphere, the spacetime mentioned in Eq.~(\ref{genspt}) can cast a shadow. Hence, the conditions for the existence of a finite size shadow without photon sphere are $\lim_{r \to 0}\frac{r}{\sqrt{A(r)}}= \alpha \,\, ,$ and $V_{eff}^{\prime}(r)<0~\forall~r,$ where $\alpha$ is a real, positive finite number and it would be the radius of the shadow. Therefore, a finite size shadow without photon sphere exists when, 
\begin{eqnarray}
\lim_{r \to 0}A(r)&=&\left(\frac{r}{\alpha}\right)^2\,\, ,\nonumber\\
\frac{rA^{\prime}(r)}{2A(r)}& < & 1 ~\forall ~ r \,\, ,
\label{con2}
\end{eqnarray}
where the second condition in the above equation is derived from $V_{eff}^{\prime}(r)<0~\forall~r$.
Using the above equation we can write, $\lim_{r \to 0}V_{eff}(r)=\left(\frac{1}{\alpha}\right)^2$. There can be many functional expressions of $A(r)$ which at $r=0$, reduce to the limiting value as stated in Eq.~(\ref{con2}). One of the simple forms of $A(r)$ can be written as, 
\begin{equation}
A(r)=\left(\frac{r}{\alpha}\right)^2\left(1+\beta r^n\right)^m\,\, ,
\end{equation}
\begin{figure}
{\includegraphics[width=90mm]{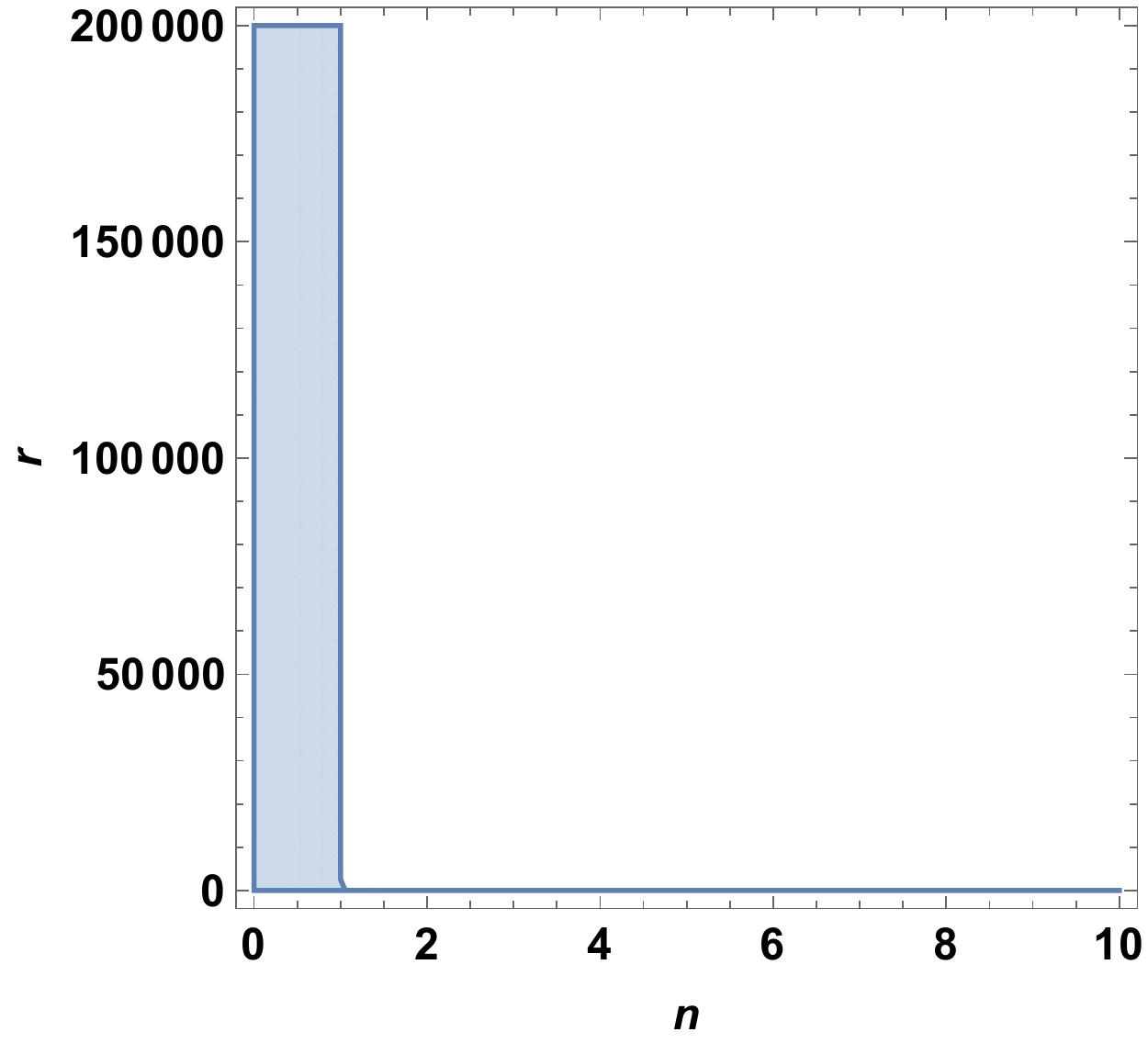}}
\caption{Figure shows the allowed range (i.e. shaded region) of $n$ where $\rho + P_{\perp}>0$ for any value of $r$. }
\label{enrgcon}
\end{figure}
where $\beta, m, n $ are constant parameters. It can be verified that in order to satisfy the condition mentioned in Eq.~(\ref{con2}), we need $n>0$, $m<0$ and $\beta>0$. Therefore, with the above expression of $A(r)$ where $n>0$, $m<0$ and $\beta>0$ the spacetime in Eq.~(\ref{genspt}) can cast shadow of radius $\alpha$, though there exist no photon sphere in that spacetime. 
%Asymptotically flatness is considered to be an important property of a physically realistic spacetime. A s There are many important spacetimes which are not asymptotically flat. 
If we consider that the spacetime in Eq.~(\ref{genspt}) is asymptotically Minkowskian, then the following additional conditions should be satisfied,
\begin{eqnarray}
n&=& -\frac{2}{m}\,\, ,\nonumber\\ 
\beta &=& \frac{1}{\alpha^n}\,\, ,
\end{eqnarray}
and we also need $\lim_{r \to \infty}B(r)=1$. Therefore, the final expression $A(r)$ for which there exist a shadow without a photon sphere and $\lim_{r \to \infty}A(r)=1$ is,
\begin{equation}
A(r)= \left(\frac{r}{\alpha}\right)^2\left(1+\left(\frac{r}{\alpha}\right)^n\right)^{-\frac{2}{n}}\,\, ,
\label{ar1}
\end{equation}
where $n$ and $\alpha$ are positive constant parameters. In \cite{Joshi2020}, we propose a spacetime which can cast shadow, though there exist no photon sphere in that spacetime. One can verify that for $n=1$, the above expression of $A(r)$ reduces to the expression of $A(r)$ of the spacetime discussed in \cite{Joshi2020} and the line element can be written as,
\begin{equation}
ds^2=-\frac{dt^2}{\left(1+\frac{\alpha}{r}\right)^2}+\left(1+\frac{\alpha}{r}\right)^2 dr^2 +r^2 d\Omega^2\,\, .
\label{joshispt}
\end{equation}
In \cite{Joshi2020}, it is shown that the above spacetime has a strong naked singularity at the center where $\alpha$ is the ADM mass of the that spacetime. 

Considering the expression of $A(r)$ given in Eq.~(\ref{ar1}), if we express $B(r)=\frac{1}{A(r)}$ then the line element in Eq.~(\ref{genspt}) can be written as,
\begin{equation}
ds^2=-\left(1+\left(\frac{\alpha}{r}\right)^n\right)^{-\frac{2}{n}}dt^2+\left(1+\left(\frac{\alpha}{r}\right)^n\right)^{\frac{2}{n}}dr^2+r^2d\Omega^2\,\, ,
\label{sptm2}
\end{equation}
where the above spacetime is asymptotically Minkowskian and it can cast shadow of radius $\alpha$ in the absence of the photon sphere. The expression of the Ricci scalar ($R$) and the Kretschmann scalar ($Kr$) of the above spacetime are,
\begin{widetext}
\begin{eqnarray}
R&=&\frac{\left(1+\left(\frac{\alpha}{r}\right)^n\right)^{-\frac{2}{n}}\left[-6+\left(1+\left(\frac{\alpha}{r}\right)^n\right)^{\frac{2}{n}}+\left(\frac{r}{\alpha}\right)^{2n}\left(-1+\left(1+\left(\frac{\alpha}{r}\right)^n\right)^{\frac{2}{n}}\right)+\left(\frac{r}{\alpha}\right)^{n}\left(-5+n+2\left(1+\left(\frac{\alpha}{r}\right)^n\right)^{\frac{2}{n}}\right)\right]}{r^2*(1+\left(\frac{r}{\alpha}\right)^n)^2}\,\, ,\\
%%%%%%%%%%%%%%%%%%%%%%%%%%%%%%%%%%%%%%%%%%%%%%%%%%%%%%%%
Kr&=&R_{\mu\nu\gamma\beta}R^{\mu\nu\gamma\beta}=\frac{4}{r^4}\left(1+\left(\frac{\alpha}{r}\right)^n\right)^{-\frac{4}{n}}\left[\frac{(2+n)^2}{(1+\left(\frac{r}{\alpha}\right)^n)^4}-\frac{2(1+n)(2+n)}{(1+\left(\frac{r}{\alpha}\right)^n)^3}+\frac{5+n(2+n)}{(1+\left(\frac{r}{\alpha}\right)^n)^2}+\left(-1+\left(1+\left(\frac{\alpha}{r}\right)^n\right)^{\frac{2}{n}}\right)^2\right]\,\, ,\nonumber\\
\end{eqnarray}
\end{widetext}
where it can be verified that both the scalar diverges at $r=0$. Therefore, the above spacetime has a strong naked singularity at the center.
However, in order to be a physically viable solution of Einstein equation the above spacetime should satisfy the weak energy condition. In Fig.~(\ref{enrgcon}), in the space of $n$ and $r$, we show the allowed region (i.e. the shaded region) for which weak energy conditions are satisfied. It can be seen that for $n>1$ the weak energy conditions are violated. Deriving the expressions of energy density ($\rho$), radial pressure ($P_r$) and azimuthal pressures ($P_{\perp}$), one can verify by  that $\rho>0$ and $\rho +P_r>0$ for all values of $r$ when $n>0$. Therefore, in that figure, we only show the allowed region where $\rho + P_{\perp}>0$.  From Fig.~(\ref{enrgcon}), one can conclude that for $n\leq 1$ the above spacetime (Eq.~(\ref{sptm2})) satisfy the weak energy conditions and it can cast shadow without photon sphere. 

The other important property of a spacetime that one need to verify is the ADM mass of the spacetime. The ADM mass of a spacetime can be written as \cite{Joshi2020},
\begin{equation}
    M_{ADM} = - \frac{1}{8\pi} \lim_{S \to \infty} \oint_S \left(\mathbb{K}-\mathbb{K}_0\right) \sqrt{\sigma}  d^2\theta\,\, ,
\label{ADM1}
\end{equation}
where $S$ is the bounded two-surface. $\mathbb{K}$ is the extrinsic curvature of the two-surface $S$ embedded in spacelike hypersurface $\Sigma$, $\mathbb{K}_0$ is the extrinsic curvature of the two-surface ($S$) embedded in flat space and the infinitesimal area of the two-surface is written as $\sqrt{\sigma}d^2\theta = r^2\sin\theta d\theta d\phi$, where $\sigma$ is the determinant of the induced metric on the two-surface $S$. The expression of the extrinsic curvature ($K$) for the spacetime written in Eq.~(\ref{sptm2}) is,
\begin{equation}
\mathbb{K}=\frac{2}{r\left(1+\frac{\alpha^n}{r^n}\right)^{\frac1n}}\,\, ,
\end{equation}
where for flat spacetime $\mathbb{K}_0 = \frac2r$. Therefore, the ADM mass of the above spacetime (Eq.~(\ref{sptm2})) can be written as,
\begin{eqnarray}
M_{ADM}=\lim_{r \to \infty}\left[\frac{r\alpha\left(1+(\frac{r}{\alpha})^n\right)^{\frac1n}-r^2}{\alpha\left(1+(\frac{r}{\alpha})^n\right)^{\frac1n}}\right]&=&\alpha\,\, , for~ n=1\nonumber\\
&=& \infty\,\, , for~ n<1\,\, .\nonumber\\
\end{eqnarray}
Therefore, the spacetime in Eq.~(\ref{sptm2}) can extend to infinity when $n=1$ and it is asymptotically flat. However, for $n<1$, though the spacetime in Eq.~(\ref{sptm2}) is asymptotically Minkowskian, the ADM mass of that spacetime is not finite. These type of spacetime is generally called as asymptotically quasi-flat spacetime \cite{Nucamendi:1996ac}. For these types of spacetime, in order to have a finite ADM mass, we need to match the spacetime at a certain matching radius with Schwarzschild spacetime. 

In \cite{Paul2020}, it is shown that the spacetime in Eq.~(\ref{joshispt}) has a strong nulllike naked singularity at the center. Therefore, one can ask whether the existence of a nulllike naked singularity is the reason behind the existence of a shadow without a photon sphere. In another way, we can ask whether a timelike naked singularity can also cast a shadow in the absence of a photon sphere. In the next section, we discuss this issue in detail.
%%%%%%%%%%%%%%%%%%%%%%%%%%%%%%%%%%%%%%%%%%%%%%%%%%%%%%%%%%%%%%%%%%%%%%%%%%%
\begin{figure}
{\includegraphics[width=90mm]{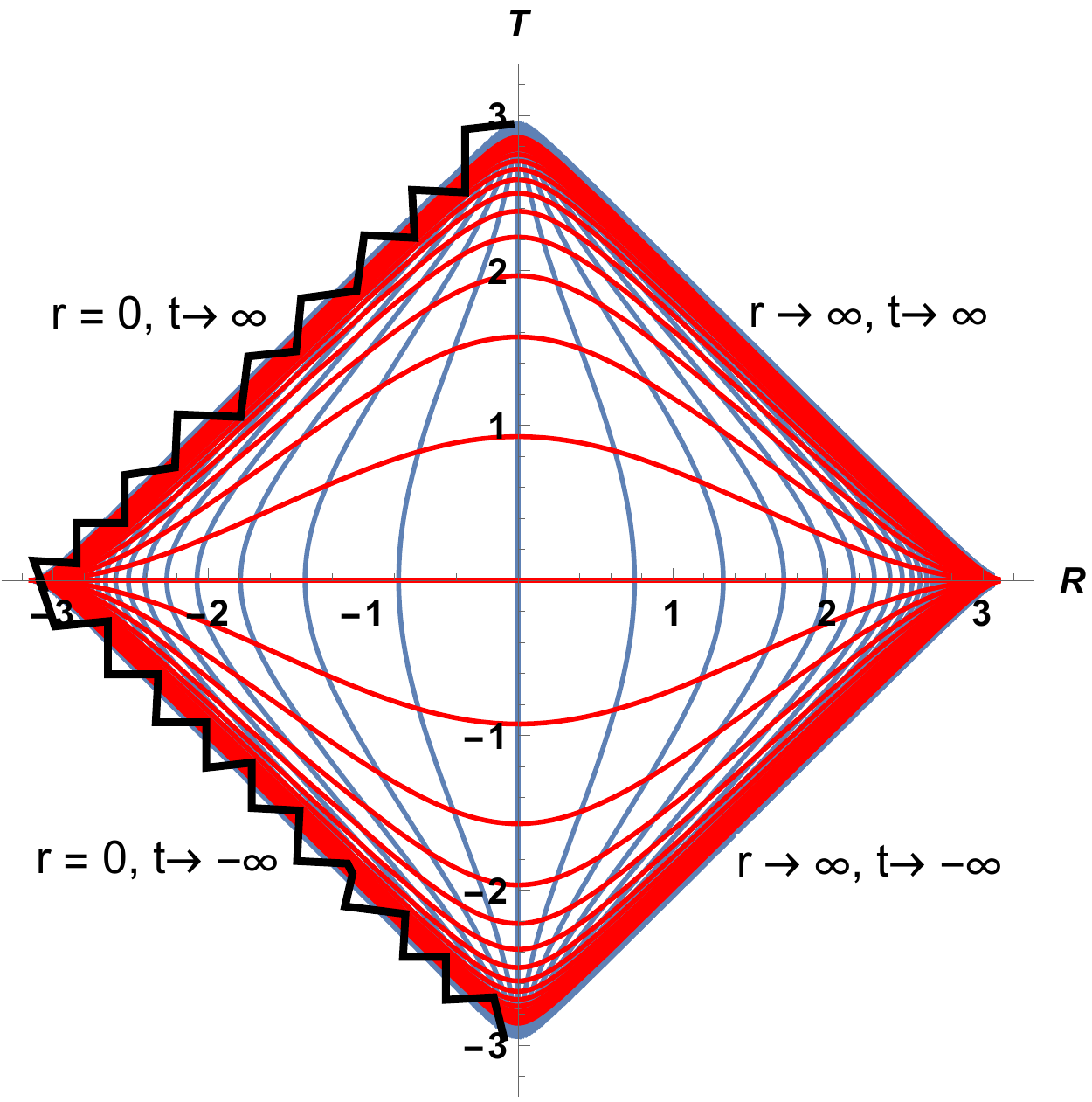}}
\caption{Figure shows Penrose diagram of the asymptotically Minkowskian spacetime written in Eq.~(\ref{sptm2}), where we consider $n=0.5$. The nulllike naked singularity at $r=0$ is shown by black zigzag lines. Along the red and blue lines time and radial distance are constant respectively.   }
\label{nulllikesing}
\vspace{1cm}
{\includegraphics[width=90mm]{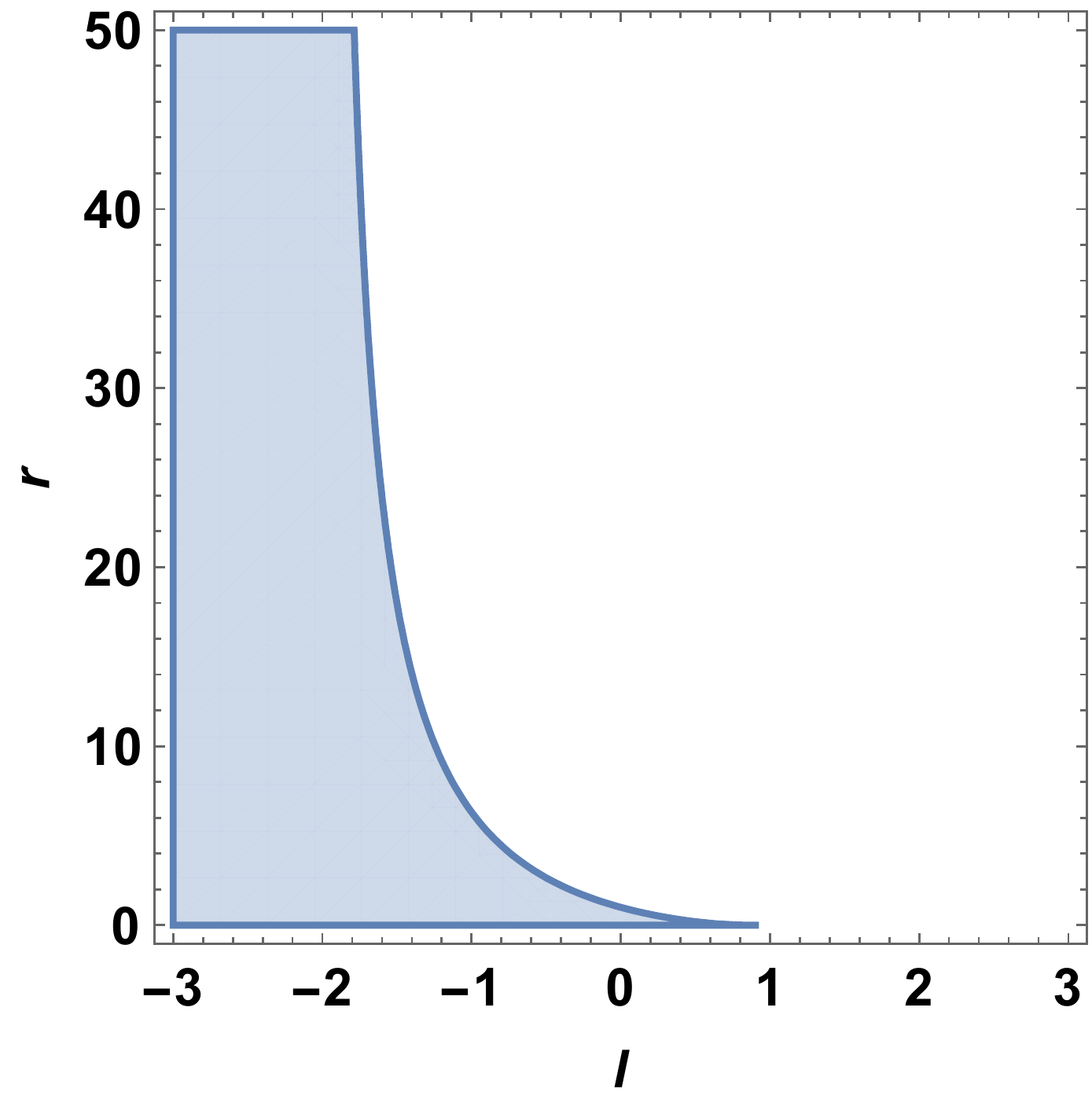}}
\caption{Figure shows the allowed range (i.e. shaded region) of $l$ where $\rho + P_{\perp}>0$ for any value of $r$.  }
\label{lvsr}
\end{figure}

%%%%%%%%%%%%%%%%%%%%%%%%%%%%%%%%%%%%%%%%%%%%%%%%%%%%%%%%%%%%%%%%%%%

\section{Possibility of shadow without photon sphere in the presence of a central timelike naked singularity}
\label{sec3}
One can verify whether a spacetime posses a timelike or nulllike naked singularity using the Penrose diagram of that spacetime. In Penrose diagram, the temporal and radial coordinates are transformed in such a way that we can describe the whole spacetime manifold in a finite size causal diagram. The coordinates $t,r$ of the general spacetime (Eq.~(\ref{genspt})) can be transformed to the coordinates $T,R$ as,
\begin{eqnarray}
T &=&\tan^{-1}\left(t+r^{\star}\right)+\tan^{-1}\left(t-r^{\star}\right)\,\, ,\\
R &=&\tan^{-1}\left(t+r^{\star}\right)-\tan^{-1}\left(t-r^{\star}\right)\,\, ,
\end{eqnarray}
where $r^{\star}= \int\sqrt{\frac{B(r)}{A(r)}}dr$. It is easy to verify that with the new $T, R$ coordinates we can describe the whole spacetime manifold (i.e. $-\infty<t<\infty$ and $0\leq r<\infty$) in a finite size diagram.

In order to investigate whether there exist a timelike or nulllike naked singularity, we need to evaluate the values of $T$ and $R$ at $r\rightarrow 0$. It can be verified that for a finite value of $t$, if $r^{\star}\rightarrow -\infty$ for $r\rightarrow 0$ then the singularity at $r=0$ is a nulllike singularity, where for finite value of $t$ and $r\rightarrow 0$ we get $R=-\pi$ and $T=0$. This is because, as we shall see, the $r=0$ hypersurface, in this case, coincides with the null-hypersurface in the Penrose diagram. On the other hand, one can define a singularity as timelike  if $r^{\star}\rightarrow 0$ for $r\rightarrow 0$. For timelike singularity, at $r=0$, we get $R=0$. Therefore, the functional form of $r^{\star}(r)$ is very important to understand the nature of singularity. Since $r^{\star}= \int\sqrt{\frac{B(r)}{A(r)}}dr$, the nature of singularity depends upon the functional form of $A(r)$ and $B(r)$ near $r=0$.
\begin{figure}
{\includegraphics[width=90mm]{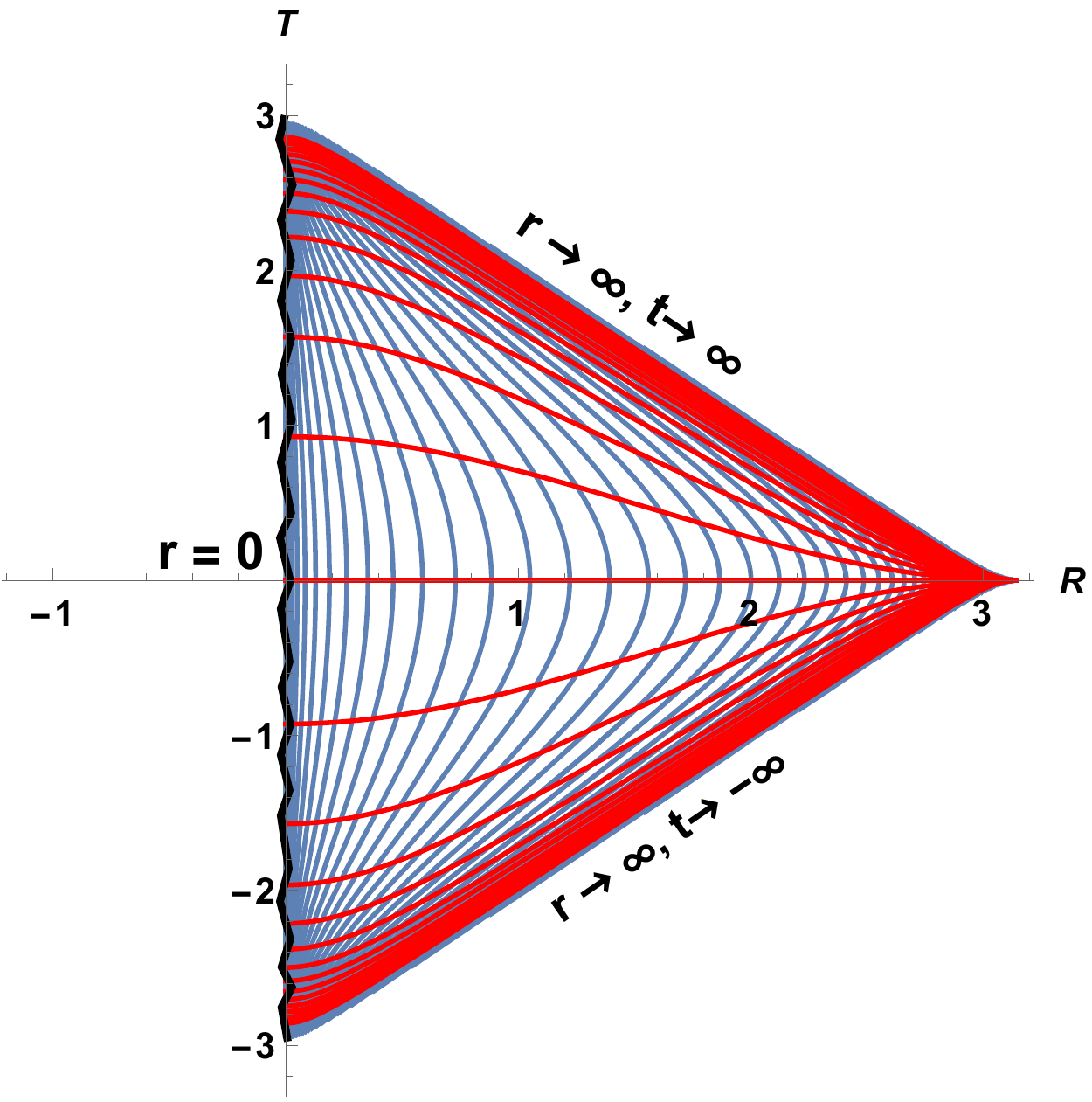}}
\caption{Figure shows Penrose diagram of the spacetime written in Eq.~(\ref{spttime1}), where we consider $n=0.5$ and $l=5$. The timelike naked singularity at $r=0$ is shown by the vertical black zigzag lines. Along the red and blue lines time and radial distance are constant respectively.   }
\label{timelikesing}
\end{figure}
From the above analysis, we can conclude that a nulllike naked singularity can cast shadow without photon sphere when the following conditions are satisfied,
\begin{eqnarray}
\lim_{r\to 0}\int\sqrt{\frac{B(r)}{A(r)}}dr &\to & -\infty\,\, ,\nonumber\\
\lim_{r\to 0}A(r) &=& \left(\frac{r}{\alpha}\right)^2\,\, ,\nonumber \\
\frac{rA^{\prime}(r)}{2A(r)}& < & 1 ~\forall ~ r \,\, ,
\label{nullcon}
\end{eqnarray}
where the first condition is for nulllike singularity and the second and third conditions are for the existence of finite size shadow without photon sphere. One can verify that the above conditions (Eq.~(\ref{nullcon})) are satisfied for the spacetime described in Eq.~(\ref{sptm2}). In Fig.~(\ref{nulllikesing}), we show the Penrose diagram for that spacetime where we consider $n=0.5$. From that figure, one can conclude that the spacetime in Eq.~(\ref{sptm2}) has a strong nulllike naked singularity at the center.
Similarly, for timelike naked singularity, we can write down the following conditions,
\begin{eqnarray}
\lim_{r\to 0}\int\sqrt{\frac{B(r)}{A(r)}}dr &= & 0\,\, ,\nonumber\\
\lim_{r\to 0}A(r) &=& \left(\frac{r}{\alpha}\right)^2\,\, ,\nonumber\\
\frac{rA^{\prime}(r)}{2A(r)}& < & 1 ~\forall ~ r \,\, .
\label{timecon}
\end{eqnarray}
If we consider $\lim_{r\to 0}B(r)=r^l$ and $A(r)$ given by Eq.~(\ref{ar1}) then near $r=0$, $r^{\star}=\frac{2}{l} r^{\frac{l}{2}}$. Therefore, in order to satisfy the condition $\lim_{r\to 0}r^{\star}=0$ (i.e. the condition for timelike singularity at $r=0$), we need $l>0$. If we consider the spacetime (Eq.~(\ref{genspt})) having timelike naked singularity at the center is asymptotically Minkowskian then we can write down one of the possible functional form of $B(r)$ as, $B(r)= \left(\frac{r}{\alpha}\right)^l\left(1+\left(\frac{r}{\alpha}\right)^k\right)^{-\frac{l}{k}}$ while retaining the form of $A(r)$ described in Eq.~(\ref{ar1}). Here, $l$ and $k$ are constant parameters which should be always positive. The positive values of $l$ and $k$ ensure that the spacetime in Eq.~(\ref{genspt}) is asymptotically Minkowskian, it has a timelike naked singularity at the center and it can cast shadow without photon sphere. Note that Eq.~(\ref{nullcon}) and Eq.~(\ref{timecon}) are the general conditions respectively for a nulllike singular and timelike singular spacetimes which can cast shadow without photon sphere. However, the conditions on $l$ and $k$ for timelike naked singularity and the conditions on $n$ for nulllike naked singularity is coming out due to the specific choice of the functional form of $A(r)$ and $B(r)$. We can write down the line element of the asymptotically Minkowskian spacetime having timelike naked singularity at the center as,
\begin{equation}
ds^2=-\left(1+\left(\frac{\alpha}{r}\right)^n\right)^{-\frac{2}{n}}dt^2+\left(1+\left(\frac{\alpha}{r}\right)^k\right)^{-\frac{l}{k}}dr^2+r^2d\Omega^2\,\, .
\label{timespt1}
\end{equation}
However, one can verify from Fig.~(\ref{lvsr}) that the weak energy conditions are invalid for all positive values of $l$ where $k>0$. In Fig.~(\ref{lvsr}), we consider $k=1$. Therefore, the timelike singular spacetime in Eq.~(\ref{timespt1}) is not physically realistic. If we remove the asymptotically Minkowskian condition from the that spacetime then we can write down it in the following simple form,
\begin{equation}
ds^2=-\left(1+\left(\frac{\alpha}{r}\right)^n\right)^{-\frac{2}{n}}dt^2+\left(\frac{r}{\alpha}\right)^{l}dr^2+r^2d\Omega^2\,\, ,
\label{spttime1}
\end{equation}
where the above spacetime can cast shadow without photon sphere and the radius of the shadow is $\alpha$. Near the center, the Ricci scalar and the Kretschmann scalar of the above spacetime can be written as,
\begin{eqnarray}
R &=& \frac{1}{r^2}\left(\frac{r}{\alpha}\right)^{-l}\left[-6+3l+2\left(\frac{r}{\alpha}\right)^l\right]\,\, ,\\
\nonumber\\
Kr &=& \frac{1}{r^4}\left(\frac{r}{\alpha}\right)^{-2l}\left[12+3l^2-8\left(\frac{r}{\alpha}\right)^l+4\left(\frac{r}{\alpha}\right)^{2l}\right]\,\, ,
\end{eqnarray}
\begin{figure*}
\centering
\subfigure[Intensity distribution in the spacetime with nulllike naked singularity]
{\includegraphics[width=82mm]{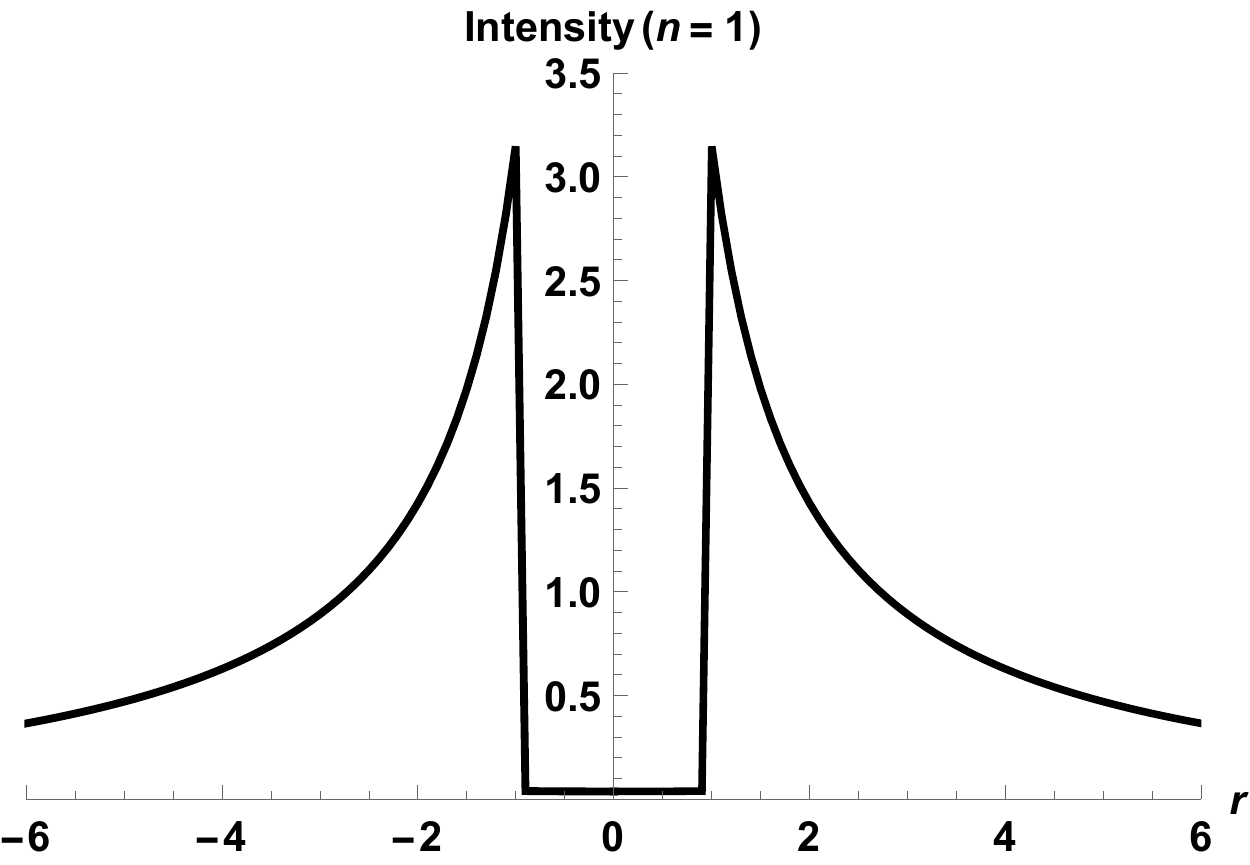}\label{fig4a}}
\hspace{0.2cm}
\subfigure[Shadow of nulllike naked singularity without photon sphere.]
{\includegraphics[width=75mm]{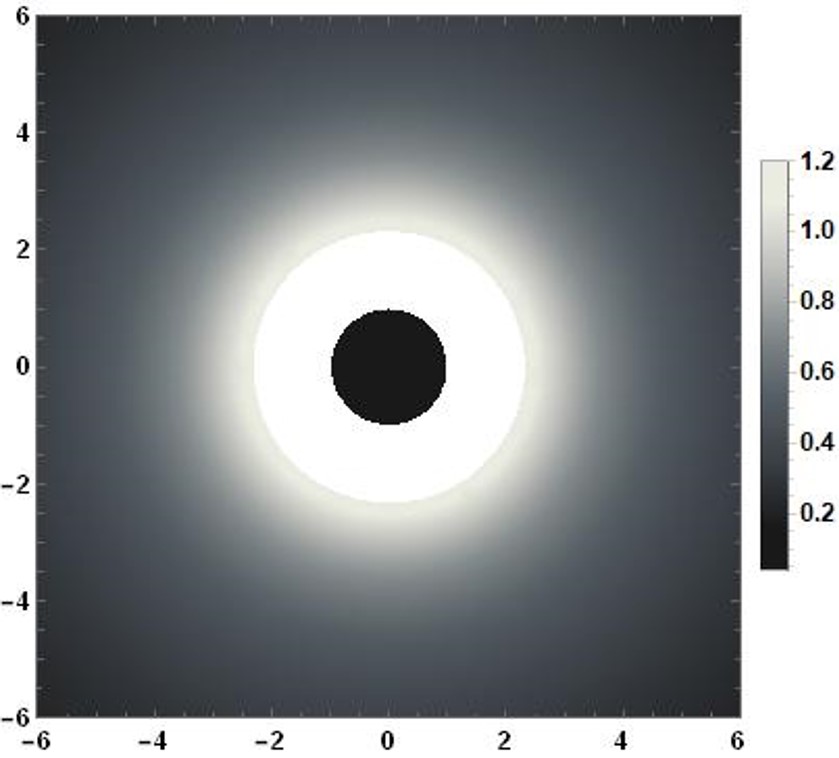}\label{fig4b}}\\
\subfigure[Intensity distribution in the spacetime with nulllike naked singularity]
{\includegraphics[width=82mm]{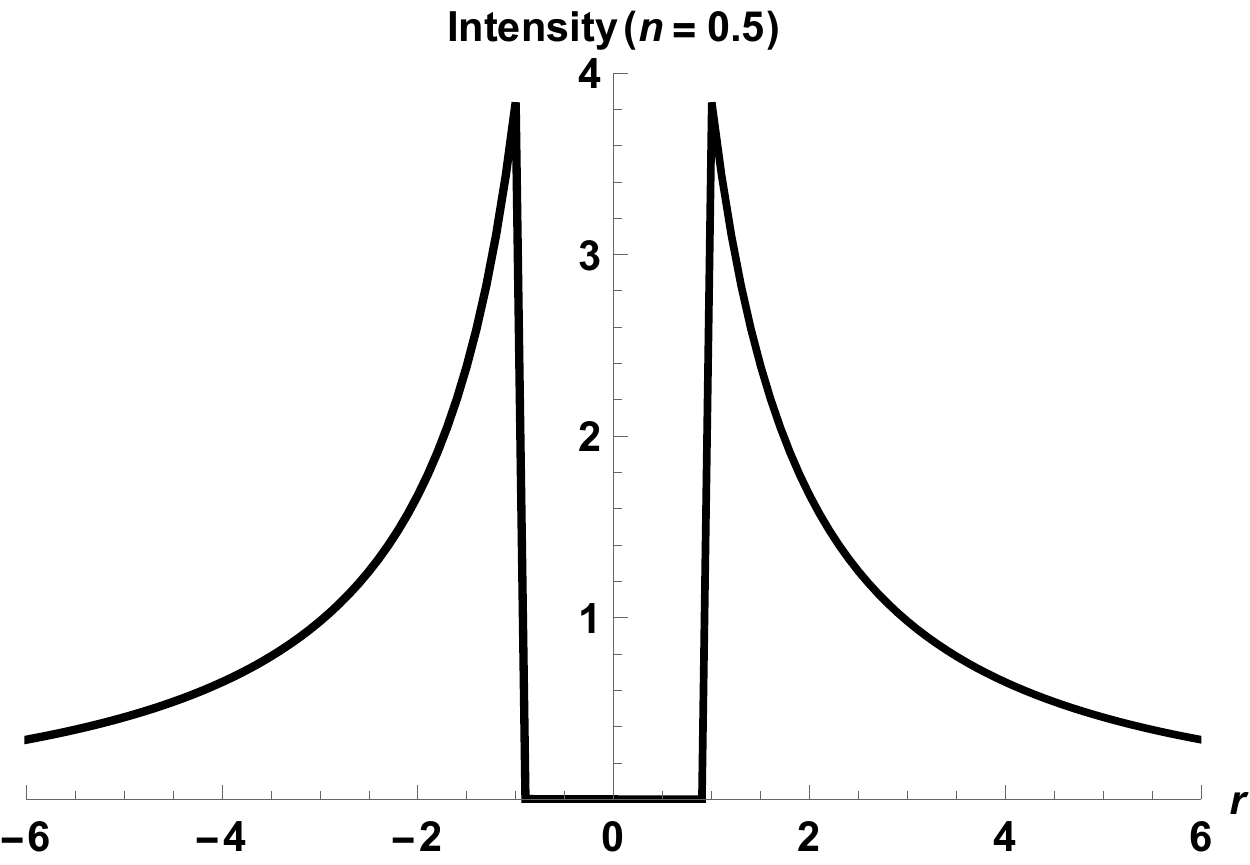}\label{fig4c}}
\hspace{0.2cm}
\subfigure[Shadow of nulllike naked singularity without photon sphere.]
{\includegraphics[width=75mm]{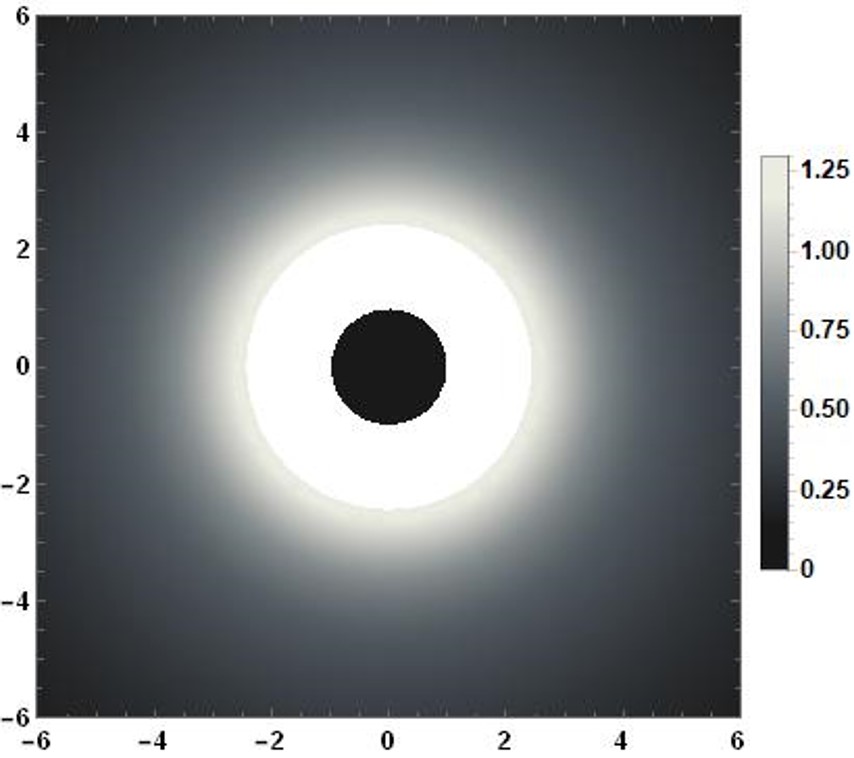}\label{fig4d}}\\
\subfigure[Intensity distribution in the spacetime with timelike naked singularity]
{\includegraphics[width=82mm]{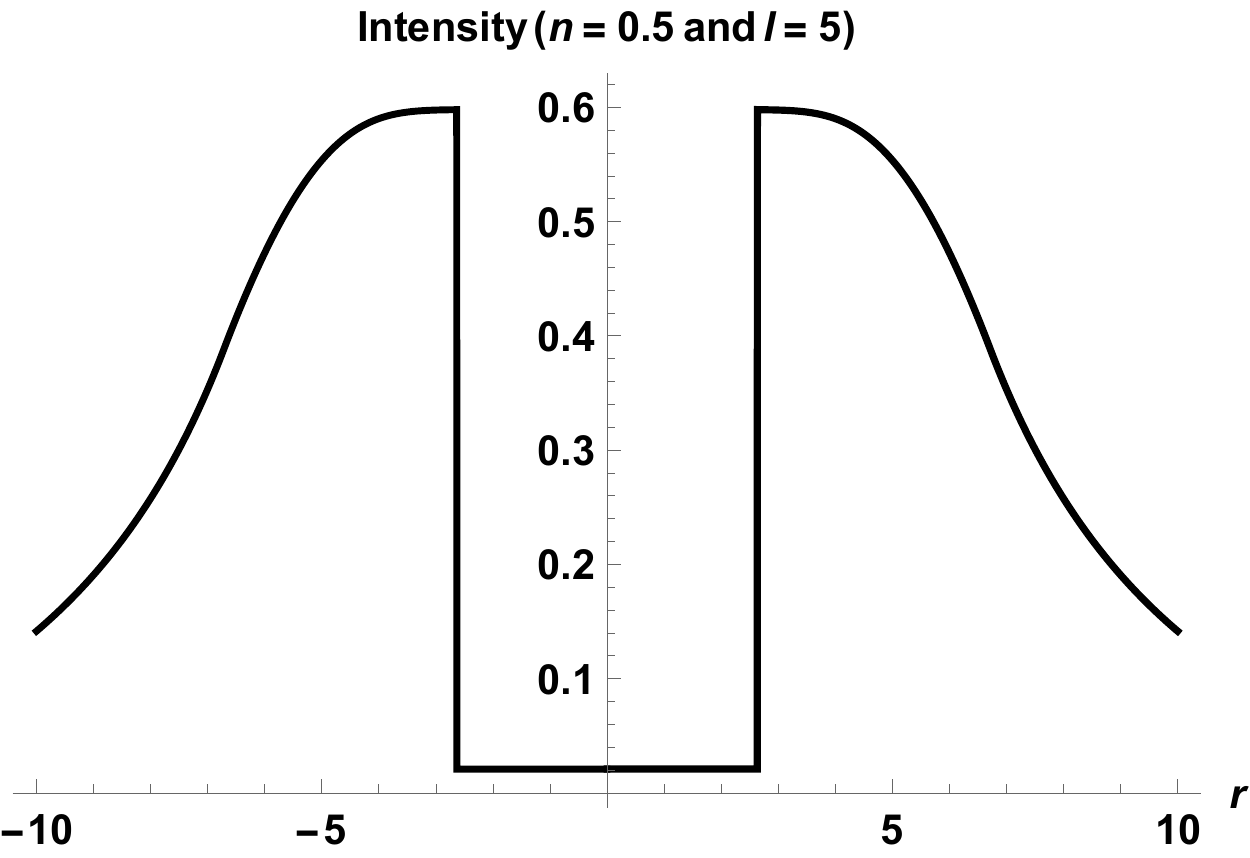}\label{fig4e}}
\hspace{0.2cm}
\subfigure[Shadow of timelike naked singularity without photon sphere.]
{\includegraphics[width=75mm]{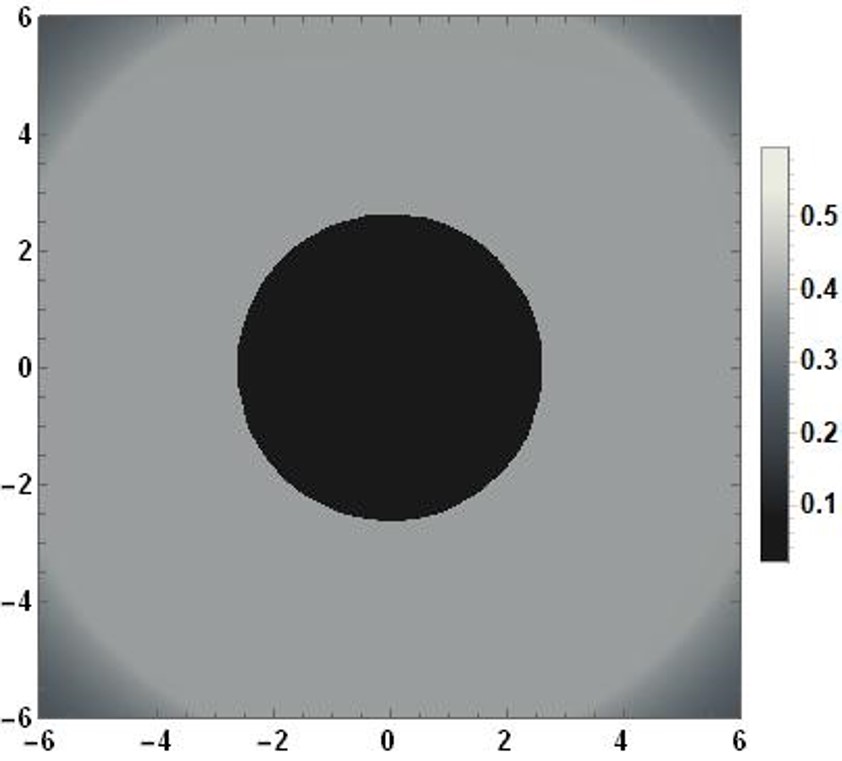}\label{fig4f}}\\
 \caption{Fig.~(\ref{fig4a}), Fig.~(\ref{fig4c}) show how the intensity varies with the impact parameter ($b$) in the presence of nulllike naked singularity and Fig.~(\ref{fig4e})  shows the same in the presence of timelike naked singularity. Fig.~(\ref{fig4b}), Fig.~(\ref{fig4d}) show the shadow cast by the nulllike naked singularities and Fig.~(\ref{fig4f}) shows the shadow cast by the timelike naked singularity in the absence of photon sphere.}
\label{fig4}
\end{figure*}
which shows the presence of a strong timelike naked singularity at the center. The Penrose diagram of the above spacetime is shown in Fig.~(\ref{timelikesing}) where $n=0.5$ and $l=5$. From the expressions of energy density ($\rho$), radial pressure ($P_r$) and tangential pressures ($P_{\perp}$) we can write,
\begin{widetext}
\begin{eqnarray}
\rho &=& \frac{1}{r^2}\left[1+(l-1)\left(\frac{r}{\alpha}\right)^{-l}\right]\,\, ,\label{rho1}\\
\rho + P_r &=& \frac{1}{r^2}\left(\frac{r}{\alpha}\right)^{-l}\left[\frac{2+l\left(1+\left(\frac{r}{\alpha}\right)^{n}\right)}{\left(1+\left(\frac{r}{\alpha}\right)^{n}\right)}\right]\,\, ,\label{rhopr1}\\
\rho + P_{\perp} &=&\frac{2\left(1+\left(\frac{r}{\alpha}\right)^{n}\right)^2+\left(\frac{r}{\alpha}\right)^{n-l}(-4+l-2n+(-2+l)\left(\frac{r}{\alpha}\right)^{n})}{2r^2\left(1+\left(\frac{r}{\alpha}\right)^{n}\right)^2}\,\, ,\label{rhopthe1}
\end{eqnarray}
\end{widetext}
Using Eq.~(\ref{rho1}) and Eq.~(\ref{rhopthe1}), one can verify that weak energy condition is valid for $l\geq 4+2n$. Therefore, for $n=0.5$ and $l\geq 5$, the weak energy conditions are valid everywhere in the above spacetime (Eq.~(\ref{spttime1})). Therefore, we conclude that like a nulllike naked singularity, a timelike naked singularity also can cast a shadow without a photon sphere.

In the next section, we discuss the images of the naked singularities in the presence of a spherically symmetric, radially freely falling accreting matter around the nulllike and timelike naked singularities.

\section{Shadow of Nulllike and timelike naked singularities in the absence of photon sphere}
\label{sec4}
Before we discuss the shadow cast by timelike naked singularity, we need to modify the timelike singular spacetime (Eq.~(\ref{spttime1})) in such a way that the spacetime can be matched smoothly with an external Schwarzschild spacetime at a certain matching radius $r_b$. This modification is necessary to get a finite Schwarzschild mass of the spacetime configuration which is internally timelike singular spacetime and externally Schwarzschild spacetime. The spacetime configuration can be written as,
\begin{widetext}
\begin{eqnarray}
ds_{int}^2&=&-(1-M_0)\left(\frac{1+\left(\frac{\alpha}{r_b}\right)^n}{1+\left(\frac{\alpha}{r}\right)^n}\right)^{\frac{2}{n}}dt^2+\frac{1}{(1-M_0)}\left(\frac{r}{r_b}\right)^{l}dr^2+r^2d\Omega^2\,\, ,\label{sptint}\\
ds_{ext}^2&=&-\left(1-\frac{M_0 r_b}{r}\right)dt^2+\frac{1}{\left(1-\frac{M_0 r_b}{r}\right)}dr^2+r^2d\Omega^2\,\, \label{sptext},
\end{eqnarray}  
\end{widetext}
where the internal spacetime (corresponding to $ds^2_{int}$) is the modified version of the timelike singular spacetime described in Eq.~(\ref{spttime1}). Here, $0<M_0<1$ and the Schwarzschild mass $M_s=\frac{M_0r_b}{2}$. As we know, in order to match two spacetimes smoothly at a certain matching radius ($r_b$), the induced metric ($h_{ab}$) and the extrinsic curvature ($K_{ab}$) corresponding to the internal and external spacetime should match at the matching radius $r_b$ \cite{Israel:1967wq,Eric,Israel66}. It is easy to verify that the induced metrics of the above two spacetimes (Eqs.~(\ref{sptint}, \ref{sptext})) match at the matching radius $r_b$. Now, the non-zero components of the extrinsic curvature of the internal and external spacetime at the matching radius are,
\begin{eqnarray}
(K_{00})_{int}&=&\frac{(1-M_0)^\frac32}{r_b\left(1+\left(\frac{r_b}{\alpha}\right)^n\right)}\, \, , (K_{00})_{ext}=\frac{M_0}{2r_b}\sqrt{1-M_0}\, ,\nonumber\\
\label{k00}\\
(K_{22})_{int}&=&(K_{22})_{ext}=r_b\sqrt{1-M_0}\,\, ,\\
(K_{33})_{int}&=&(K_{33})_{ext}=r_b\sqrt{1-M_0} \sin^2(\theta)\,\, .
\end{eqnarray} 
Therefore, in order to smoothly match the internal and external spacetimes (Eqs.~(\ref{sptint}), (\ref{sptext})), the matching radius should be,
\begin{equation}
r_b = \alpha\left[\frac{2}{M_0}\left(1-\frac{3M_0}{2}\right)\right]^\frac1n\,\, ,
\label{rb}
\end{equation}
which can be obtained by using Eq.~(\ref{k00}). At the above mentioned matching radius the internal timelike singular spacetime (Eq.~(\ref{sptint})) smoothly matches with the external Schwarzschild spacetime (Eq.~(\ref{sptext})). Now, we are ready to simulate the shadow cast by the spacetime configuration which has a finite Schwarzschild mass $M_s$. To do so, we follow \citep{Shaikh:2018lcc} where the authors consider radiation emitted from a radially freely falling spherically symmetric accereting matter and produce the intensity map. However, we do not put any mathematical detail here as it can be found in that paper.

%%%%%%%%%%%%%%%%%%%%%%%%%%%%%%%%%%%%%%%%%%%%%%%%%%%55
\begin{widetext}
\centering
   \begin{table*}
    \centering
    \caption{\textbf{Mass of Sagittarius A* and its possible shadow size in the presence of Schwarzschild Black hole, nulllike and timelike naked singularities}}
    \begin{tabular}{l c c}
        \hline\hline
Object   &   Mass   &   Shadow (diameter) \\
     &   ($M_{\odot}$)   &   ($\mu arc sec$) \\
\hline
Black hole  &    $(4.31\pm 0.38)\times10^6$   &  $56\pm8$ \\
& &\\
Nulllike naked singularity  &     $(4.31\pm 0.38)\times10^6$  & $10\pm 1.54$ \\
($n=1$) & & \\
 & &\\
Timelike naked singularity  &    $(4.31\pm 0.38)\times10^6$  & $12.48\pm 1.784$ \\
($n=0.8, l=6, M_0=0.233$) & & \\
 & &\\ 
Timelike naked singularity  &    $(4.31\pm 0.38)\times10^6$  & $28\pm 4$ \\
($n=0.5, l=5, M_0=0.377$) & & \\
 & &\\
Timelike naked singularity  &    $(4.31\pm 0.38)\times10^6$  & $85.03\pm 12.143$ \\
($n=0.35, l=5, M_0=0.423$) & & \\
\hline
    \end{tabular}
    \label{tabl1}
\end{table*}
\end{widetext}
%%%%%%%%%%%%%%%%%%%%%%%%%%%%%%%%%%%%%%%%%%%%%%%%%%%%%
As we discussed before, for a nulllike naked singularity, the spacetime in Eq.~(\ref{sptm2}) can cast a shadow of radius $\alpha$ in the absence of a photon sphere, where for $n=1$ the ADM mass has a finite value $\alpha$ and for $n<1$ the ADM mass is infinite though the spacetime is asymptotically Minkowskian. The timelike singular spacetime in Eq.~(\ref{spttime1}) is not asymptotically Minkowskian and also it has infinite ADM mass. Therefore, to keep the mass finite, we modify the spacetime in such a way that it can match with an external Schwarzschild spacetime smoothly at a finite radius (see Eqs.~(\ref{sptint},\ref{sptext})). This spacetime configuration can cast a shadow of finite radius in the absence of a photon sphere and it has also a finite Schwarzschild mass (i.e. $M_s=\frac{M_0 r_b}{2}$). The radius of the shadow ($b_s$) cast by the spacetime configuration is,
\begin{equation}
b_s =\frac{\alpha}{\sqrt{1-M_0}}\left(1+\left(\frac{\alpha}{r_b}\right)^n\right)^{-\frac{1}{n}}\,\, .
\label{bs}
\end{equation}
In Fig.~(\ref{fig4e}), we show how the intensity of light emitted by the radially freely falling spherically symmetric accereting matter varies with impact parameter in the presence of central timelike naked singularity, where we take $n=0.5, ~l=5, ~\alpha=1, ~M_s=1$. Using those parameters' values and using Eqs.~(\ref{rb}, \ref{bs}), we get $M_0=0.377$ and $r_b=5.303$, $b_s = 2.635$. Therefore, with the above mentioned parameters' values the spacetime configuration (Eq.~(\ref{sptint},\ref{sptext})) having a central timelike naked singularity cast shadow of radius $b_s=2.635$ in the absence of photon sphere. In Fig.~(\ref{fig4f}), we simulate the shadow cast by the spacetime configuration. In Fig.~(\ref{fig4a}) and Fig.~(\ref{fig4c}), we show the intensity variation in the nulllike singular spacetime (Eq.~(\ref{sptm2})) for $n=1$ and $n=0.5$ respectively and in Figs.~(\ref{fig4b},\ref{fig4d}), we show the corresponding simulated images of shadow in the asymptotic observer sky. In the both cases, the nulllike singular spacetime cast shadow of radius $\alpha =1$. However, from the Eq.~(\ref{bs}), in can be seen that the radius of the shadow of timelike naked singularity changes with $M_0, n, a$. As we mentioned earlier, the Schwarzschild mass $M_s = \frac{M_0 r_b}{2}$ and $r_b$ depends upon $M_0, n, a$ (Eq.~(\ref{rb})). Therefore, if we fix the values of $M_s$ and $\alpha$ to unity then for different values of $n$, we get different values of $M_0, r_b$ and $b_s$. It can be verified that the shadow radius of the timelike naked singularity can be less than or greater than $3\sqrt3$ which is the radius of the shadow cast by the Schwarzschild black hole. 

From the stellar motions and other physical phenomenon around the compact object Sgr-A* located at the center of our galaxy, the estimated mass of Sgr-A* is $4.3\times 10^6 M_{\odot}$ with an error of $\pm 0.38\times 10^6 M_{\odot}$ \cite{Bambi:2013nla}. If the compact object is a Schwarzschild black hole then the angular diameter of the shadow observed from earth would be $56\pm 8 ~\mu arcsec$\cite{Bambi:2013nla}. However, that diameter of the shadow would be different in the presence of nulllike and timelike naked singularities which can cast shadow without photon sphere.  In Table(\ref{tabl1}), we show the diameter of the shadow cast by nulllike and timelike naked singularities for the fixed mass $(4.3\pm 0.38)\times 10^6 M_{\odot}$. It can be seen from the table that in the presence of timelike naked singularity, the angular diameter of the shadow can be greater than or less than the angular diameter of the shadow cast by a Schwarzschild black hole. It can be verified that for $n=0.35, M_0= 0.423$, the shadow size would be $1.5$ times greater than the shadow in Schwarzschild spacetime.   On the other hand, in the presence of nulllike naked singularity, the shadow size is always shorter than the shadow of the Schwarzschild black hole. However, as it was discussed before, both the timelike and nulllike naked singularities (Eq.~(\ref{sptm2}, \ref{sptint})) cast shadow in the absence of photon sphere.
%%%%%%%%%%%%%%%%%%%%%%%%%%%%%%%%%%%%%%%%%%%%%%%%%%%%%%%%%%%%%%%%%%%%%%%%%%%%%%%%%%%%%%%%%%%%%%%%%%%%%%%%%%%

\section{Conclusion}
\label{sec5}
Few concluding important points regarding the shadow of nulllike and timelike naked singularities in the absence of photon sphere are discussed below:

\begin{itemize}
\item In this paper, we derive a general condition (Eq.~(\ref{con2})) that a spacetime should satisfy in order to cast a shadow without a photon sphere. Using the condition, we derive a class of spacetimes (Eq.~(\ref{sptm2})) which is asymptotically Minkowskian and which has a nulllike strong naked singularity at the center. We show that the radius ($\alpha$) of the shadow of nulllike singularity is independent of parameter $n$ and therefore, for the whole class of spacetimes, the shadow size remains unchanged.

\item We investigate whether timelike naked singularity also can cast shadow without a photon sphere. Using the Penrose diagram, we derive a general condition for both the nulllike (Eq.~(\ref{nullcon})) and timelike (Eq.~(\ref{timecon})) naked singularities to be able to cast a shadow in the absence of a photon sphere. We derive a class of timelike (Eq.~\ref{spttime1}) singular spacetimes from those conditions. However, since the class of the timelike naked singular spacetimes is not asymptotically Minkowskian, we modify the spacetime and propose a spacetime configuration (Eqs.~(\ref{sptint},\ref{sptext})) where the modified timelike singular spacetime smoothly matches with the external Schwarzschild spacetime at a certain matching radius (Eq.~(\ref{rb})). We show that the spacetime configuration casts shadow, though there exists no photon sphere.

\item In order to discuss our results in a physical context, we fixed the masses of the nulllike and timelike singular spacetimes to the estimated mass of the central compact object Sgr-A* of our Milky-way galaxy. From Table(\ref{tabl1}), it can be seen that the timelike naked singularity can cast shadow of diameter greater or less than the diameter of shadow cast by a Schwarzschild black hole with same Schwarzschild mass ($M_s$). On the other hand, a nulllike naked singularity always casts a shadow of smaller radius than the radius of shadow of a Schwarzschild black hole. Therefore, any observational result which shows the size of the shadow of a compact object, which  is greater than the expected shadow size of a Schwarzschild black hole may imply the existence of a timelike naked singularity at the center of the compact object.

\item Since the nulllike naked singularity (Eq.~(\ref{sptm2})) and the timelike naked 
singularity  (Eq.~(\ref{sptint},\ref{sptext})) cast shadows in the absence of photon spheres, any quantum gravity effects near the singularity may cause possible observable deformations in the shadow size.  
\end{itemize}

As we discussed above, recent observations of the shadow of M87 galactic center and the upcoming observational results of the shadow of Milky-way galactic center (Sgr-A*) can give us the information of the causal structure of these regions. In this context, the results reported here in this paper seem to be intriguing and worth exploring further. 

%%%%%%%%%%%%%%%%%%%%%%%%%%%%%%
%%%%%% References
%%%%%%%%%%%%%%%%%%%%%%%%%%%%%%
%%%%%%%%%%%%%%%%%%%%%%%%%%%%%%
%https://cdn.journals.aps.org/files/styleguide-pr.pdf


\begin{thebibliography}{99}

\bibitem{Akiyama:2019fyp} 
  K. Akiyama {\it et al.} [Event Horizon Telescope Collaboration],
  %``First M87 Event Horizon Telescope Results. V. Physical Origin of the Asymmetric Ring,''
  \href{https://iopscience.iop.org/article/10.3847/2041-8213/ab0f43/meta}{  Astrophys. J.  {\bf 875}, no. 1, L5 (2019).}

\bibitem{M87} 
{Gravity Collaboration} and {Abuter} {\it et al.}
\href{https://ui.adsabs.harvard.edu/abs/2018A}{app. {\bf 618} L10 }

\bibitem{Eisenhauer:2005cv} 
  F. Eisenhauer {\it et al.},
 ``SINFONI in the Galactic Center: Young stars and IR flares in the central light month,''
\href{https://iopscience.iop.org/article/10.1086/430667/meta}{  Astrophys. J.  {\bf 628}, 246 (2005).}

\bibitem{center1}
S. Gillessen {\it et al.},
\href{https://iopscience.iop.org/article/10.3847/1538-4357/aa5c41/meta}{Astrophys. J.  {\bf 837}, 30 (2017). }


\bibitem{Joshi:1993zg}
P.~S.~Joshi and I.~H.~Dwivedi,
%``Naked singularities in spherically symmetric inhomogeneous Tolman-Bondi dust cloud collapse,''
\href{https://journals.aps.org/prd/abstract/10.1103/PhysRevD.47.5357}{Phys. Rev. D \textbf{47}, 5357-5369 (1993).}


\bibitem{JMN11} 
P. S. Joshi, D. Malafarina, and R. Narayan, 
%``Equilibrium configurations from gravitational collapse,''
\href{http://iopscience.iop.org/article/10.1088/0264-9381/28/23/235018/meta}{Class. Quantum Grav. {\bf 28}, 235018 (2011).}

\bibitem{Mosani1}
K.~Mosani, D.~Dey and P.~S.~Joshi,
%``Strong curvature naked singularities in spherically symmetric perfect fluid collapse,''
\href{https://journals.aps.org/prd/abstract/10.1103/PhysRevD.101.044052}{Phys. Rev. D \textbf{101}, no.4, 044052 (2020).}

\bibitem{Mosani2}
K.~Mosani, D.~Dey and P.~S.~Joshi,
%``Strong curvature naked singularities in spherically symmetric perfect fluid collapse,''
\href{https://journals.aps.org/prd/abstract/10.1103/PhysRevD.101.044052}{Phys. Rev. D \textbf{102}, 044037 (2020).}

\bibitem{Dafermos:2017dbw}
M.~Dafermos and J.~Luk,
%``The interior of dynamical vacuum black holes I: The $C^0$-stability of the Kerr Cauchy horizon,''
\href{https://arxiv.org/abs/1710.01722}{arXiv:1710.01722.}

\bibitem{Dey:2019fja} 
  D.~Dey, P.~Kocherlakota and P.~S.~Joshi,
  %``A General Relativistic Approach to Small-Scale Structure Formation,''
\href{https://journals.aps.org/prd/abstract/10.1103/PhysRevD.101.043005}{Phys. Rev. D \textbf{101}, no.4, 043005 (2020).}

\bibitem{Penrose} R. Penrose, Riv. Nuovo Cimento Soc. Ital. Fis. \textbf{1}, 252 (1969).


\bibitem{Shaikh:2019hbm} 
  R.~Shaikh and P.~S.~Joshi,
\href{https://iopscience.iop.org/article/10.1088/1475-7516/2019/10/064/pdf}{Journal of Cosmology and Astroparticle Physics {\bf 10}, 064, 2019.}  
  %``Can we distinguish black holes from naked singularities by the images of their accretion disks?,''


\bibitem{Gralla:2019xty} 
  S.~E.~Gralla, D.~E.~Holz and R.~M.~Wald,
  %``Black Hole Shadows, Photon Rings, and Lensing Rings,''
\href{https://journals.aps.org/prd/abstract/10.1103/PhysRevD.100.024018}{Phys.\ Rev.\ D {\bf 100}, no. 2, 024018 (2019).}

\bibitem{Abdikamalov:2019ztb} 
  A.~B.~Abdikamalov, A.~A.~Abdujabbarov, D.~Ayzenberg, D.~Malafarina, C.~Bambi and B.~Ahmedov,
  %``Black hole mimicker hiding in the shadow: Optical properties of the $\gamma$ metric,''
\href{https://journals.aps.org/prd/abstract/10.1103/PhysRevD.100.024014}{Phys.\ Rev.\ D {\bf 100}, no. 2, 024014 (2019).}

\bibitem{Yan:2019etp} 
  H.~Yan,
  %``Influence of a plasma on the observational signature of a high-spin Kerr black hole,''
\href{https://journals.aps.org/prd/abstract/10.1103/PhysRevD.99.084050}{Phys.\ Rev.\ D {\bf 99}, no. 8, 084050 (2019).}

\bibitem{Vagnozzi:2019apd} 
  S.~Vagnozzi and L.~Visinelli,
  %``Hunting for extra dimensions in the shadow of M87*,''
\href{https://journals.aps.org/prd/abstract/10.1103/PhysRevD.100.024020}{Phys.\ Rev.\ D {\bf 100}, no. 2, 024020 (2019).}

\bibitem{Gyulchev:2019tvk} 
  G.~Gyulchev, P.~Nedkova, T.~Vetsov and S.~Yazadjiev,
  %``Image of the Janis-Newman-Winicour naked singularity with a thin accretion disk,''
\href{https://journals.aps.org/prd/abstract/10.1103/PhysRevD.100.024055}{Phys.\ Rev.\ D {\bf 100}, no. 2, 024055 (2019).}

\bibitem{Shaikh:2019fpu} 
  R.~Shaikh,
  %``Black hole shadow in a general rotating spacetime obtained through Newman-Janis algorithm,''
\href{https://journals.aps.org/prd/abstract/10.1103/PhysRevD.100.024028}{Phys.\ Rev.\ D {\bf 100}, no. 2, 024028 (2019).}


\bibitem{Dey:2013yga} 
  D. Dey, K. Bhattacharya and T. Sarkar,
  %``Astrophysics of Bertrand spacetimes,''
\href{https://journals.aps.org/prd/abstract/10.1103/PhysRevD.88.083532}{Phys.\ Rev.\ D {\bf 88}, 083532 (2013).} 

\bibitem{Dey+15} 
D. Dey, K. Bhattacharya and T. Sarkar,
%``Galactic Dark Matter and Bertrand spacetimes,''
\href{https://journals.aps.org/prd/abstract/10.1103/PhysRevD.87.103505}{Phys. Rev. D {\bf 87}, 10, 103505 (2013).}

   
   \bibitem{Martinez:2019nor} 
  C.~Martínez, N.~Parra, N.~Valdés and J.~Zanelli,
  %``Geodesic Structure of Naked Singularities in AdS$_3$ Spacetime,''
  \href{https://arxiv.org/abs/1902.00145}{arXiv:1902.00145 [hep-th],(2019)}  

 \bibitem{Eva}
 %Complete Analytic Solution of the Geodesic Equation in d–(Anti-)de Sitter Spacetimes
Eva Hackmann and Claus Lämmerzahl
\href{https://journals.aps.org/prl/abstract/10.1103/PhysRevLett.100.171101}{Phys. Rev. Lett. 100, 171101 – Published 2 May 2008}
 
 \bibitem{Eva1}
 %Analytic solutions of the geodesic equation in axially symmetric space-times
E. Hackmann, V. Kagramanova2, J. Kunz2 and C. Lämmerzahl1
\href{https://iopscience.iop.org/article/10.1209/0295-5075/88/30008}{Published 13 November 2009 • Europhysics} 

\bibitem{Eva2}
%Motion of spinning test bodies in Kerr spacetime
Eva Hackmann, Claus Lämmerzahl, Yuri N. Obukhov, Dirk Puetzfeld, and Isabell Schaffer
\href{https://journals.aps.org/prd/abstract/10.1103/PhysRevD.90.064035}{Phys. Rev. D 90, 064035 – Published 22 September 2014}
   


\bibitem{tsirulev}
%Bound orbits near scalar field naked singularities
I. M. PotashovJu. V. TchemarinaA. N. Tsirulev
\href{https://link.springer.com/article/10.1140%2Fepjc%2Fs10052-019-7192-7}{The European Physical Journal C, August 2019, 79:709}


  
\bibitem{Joshi:2019rdo} 
  A.~B.~Joshi, P.~Bambhaniya, D.~Dey and P.~S.~Joshi,
  %``Timelike Geodesics in Naked Singularity and Black Hole Spacetimes II,''
  arXiv:1909.08873 [gr-qc].  



  

\bibitem{Bambh} 
  P.~Bambhaniya, A.~B.~Joshi, D.~Dey and P.~S.~Joshi,
  %``Timelike geodesics in Naked Singularity and Black Hole Spacetimes,''
\href{https://journals.aps.org/prd/abstract/10.1103/PhysRevD.100.124020}{Phys.\ Rev.\ D {\bf 100}, no. 12, 124020 (2019.)}



\bibitem{Bhattacharya:2017chr}
K.~Bhattacharya, D.~Dey, A.~Mazumdar and T.~Sarkar,
%``New class of naked singularities and their observational signatures,''
\href{https://journals.aps.org/prd/abstract/10.1103/PhysRevD.101.043005}{Phys. Rev. D \textbf{101}, no.4, 043005 (2020).}

\bibitem{Bambhaniya:2019pbr}
P.~Bambhaniya, A.~B.~Joshi, D.~Dey and P.~S.~Joshi,
%``Timelike geodesics in Naked Singularity and Black Hole Spacetimes,''
\href{https://journals.aps.org/prd/abstract/10.1103/PhysRevD.100.124020}{Phys. Rev. D \textbf{100}, no.12, 124020 (2019).}

\bibitem{Dey:2019fpv} 
  D.~Dey, P.~S.~Joshi, A.~Joshi and P.~Bambhaniya,
  %``Towards an observational test of black hole versus naked singularity at the galactic center,''
\href{https://www.worldscientific.com/doi/abs/10.1142/S0218271819300246}{Int.\ J.\ Mod.\ Phys.\ D {\bf 28}, no. 14, 1930024 (2019).}


\bibitem{Bam2020}
P.~Bambhaniya, D.~N.~Solanki, D.~Dey, A.~B.~Joshi, P.~S.~Joshi and V.~Patel,
%``Precession of timelike bound orbits in Kerr spacetime,''
\href{https://arxiv.org/abs/2007.12086}{arXiv:2007.12086 [gr-qc].}

\bibitem{Shaikh:2018lcc} 
  R.~Shaikh, P.~Kocherlakota, R.~Narayan and P.~S.~Joshi,
  %``Shadows of spherically symmetric black holes and naked singularities,''
\href{https://academic.oup.com/mnras/article-abstract/482/1/52/5113467?redirectedFrom=fulltext}{ Mon.\ Not.\ Roy.\ Astron.\ Soc.\  {\bf 482}, 52 (2019).}

\bibitem{Joshi2020}
A.~B.~Joshi, D.~Dey, P.~S.~Joshi and P.~Bambhaniya,
%``Shadow of a Naked Singularity without Photon Sphere,''
\href{https://journals.aps.org/prd/abstract/10.1103/PhysRevD.102.024022}{Phys. Rev. D \textbf{102}, no.2, 024022 (2020).}

\bibitem{Paul2020}
S.~Paul,
%``Strong gravitational lensing by a strongly naked null singularity,''
\href{https://arxiv.org/abs/2007.05509}{arXiv:2007.05509 [gr-qc].}

\bibitem{Dey:2020haf}
D.~Dey, R.~Shaikh and P.~S.~Joshi,
%``Perihelion Precession and Shadows near Blackholes and Naked Singularities,''
\href{https://journals.aps.org/prd/abstract/10.1103/PhysRevD.102.044042}{Phys. Rev. D \textbf{102}, no.4, 044042 (2020).}


\bibitem{Israel:1967wq} 
  W.~Israel,
  %``Event horizons in static vacuum space-times,''
\href{https://journals.aps.org/pr/abstract/10.1103/PhysRev.164.1776}{Phys.\ Rev.\  {\bf 164}, 1776 (1967).}



   
%\bibitem{HawkingEllis73} 
%Hawking S. W., Ellis G. F. R., 
%\textit{The Large Scale Structure of Spacetime}, (Cambridge University Press, Cambridge, 1973)

\bibitem{Eric} 
  E.~Poisson,
  %``Timelike geodesics in Naked Singularity and Black Hole Spacetimes,''
\href{https://www.cambridge.org/core/books/relativists-toolkit/DA7ED68B971708A0F782257F948981E7}{Cambridge University Press. (2004) doi:10.1017/CBO9780511606601 
}


\bibitem{Israel66}
W. Israel, 
\href{https://link.springer.com/article/10.1007/BF02710419}{Nuovo Cimento B {\bf 44}, 1 (1966).}


\bibitem{Bambi:2013nla} 
  C.~Bambi,
  %``Can the supermassive objects at the centers of galaxies be traversable wormholes? The first test of strong gravity for mm/sub-mm very long baseline interferometry facilities,''
\href{https://journals.aps.org/prd/abstract/10.1103/PhysRevD.87.107501}{Phys.\ Rev.\ D {\bf 87}, 107501 (2013).}
 

\bibitem{Nucamendi:1996ac}
U.~Nucamendi and D.~Sudarsky,
%``Quasiasymptotically flat space-times and their ADM mass,''
\href{https://iopscience.iop.org/article/10.1088/0264-9381/14/5/031}{Class. Quant. Grav. \textbf{14}, 1309-1327 (1997).}

\end{thebibliography}
\end{document}